\documentclass[5p,twocolumn]{elsarticle}
\usepackage{amssymb,epsfig,graphicx,amssymb,subeqnarray, multirow, pstricks,colortbl, bm}
\usepackage{units,tabularx,array,booktabs,booktabs,algorithm, algorithmic}
\usepackage[cmex10]{amsmath}
\usepackage[tight,footnotesize]{subfigure}  
\usepackage{soul, verbatim}
\newcommand{\be}{\begin{equation}}
\newcommand{\ee}{\end{equation}}
\newcommand{\bean}{\begin{eqnarray}}
\newcommand{\eean}{\end{eqnarray}}
\newcommand{\bea}{\begin{eqnarray*}}
\newcommand{\eea}{\end{eqnarray*}}
\newcommand{\bfig}{\begin{figure}}
\newcommand{\efig}{\end{figure}}
\newcommand{\ba}{\begin{array}}
\newcommand{\ea}{\end{array}}

\newcommand{\nind}{\newline\indent}

\newcommand{\etal}{\emph{et al. }}
\newcommand{\etalns}{\emph{et al.}}

\PassOptionsToPackage{hyphens}{url}\usepackage{hyperref}
\usepackage[]{xcolor}
\usepackage{paralist}
\journal{arxiv.org}

\begin{document}

\begin{frontmatter}

\title{Using path signatures to predict a diagnosis of Alzheimer's disease.}

\author{P.J. Moore\corref{cor1}\fnref{fn1}}
\ead{moorep@maths.ox.ac.uk}
\cortext[cor1]{Corresponding author}
\fntext[fn1]{Postdoc, Mathematical Institute, University of Oxford. } %Funded by the MRC Dementias Platform UK (DPUK) .}

\author{J. Gallacher\fnref{fn2}}
\fntext[fn2]{Professor, Department of Psychiatry, University of Oxford.}

\author{T.J. Lyons \fnref{fn3}}
\fntext[fn3]{Professor, Mathematical Institute, University of Oxford.}

\author{\\for the Alzheimer’s Disease Neuroimaging Initiative \fnref{fn4}}
\fntext[fn4]{Data used in preparation of this article were obtained from the Alzheimer’s Disease
Neuroimaging Initiative (ADNI) database (adni.loni.usc.edu). As such, the investigators
within the ADNI contributed to the design and implementation of ADNI and/or provided data
but did not participate in analysis or writing of this report. A complete listing of ADNI
investigators can be found at:\url{http://adni.loni.usc.edu/}.}

\address{University of Oxford}

%\author{P.J. Moore}

\begin{abstract}
The path signature is a means of feature generation that can encode nonlinear interactions in the data as well as the usual linear features.  It can distinguish the ordering of time-sequenced changes: for example whether or not the hippocampus shrinks fast, then slowly or the converse.  It provides interpretable features and its output is a fixed length vector irrespective of the number of input points so it can encode longitudinal data of varying length and with missing data points. In this paper we demonstrate the path signature in providing features to distinguish a set of people with Alzheimer's disease from a matched set of healthy individuals.  The data used are volume measurements of the whole brain, ventricles and hippocampus from the Alzheimer's Disease Neuroimaging Initiative (ADNI).  The path signature method is shown to be a useful tool for the processing of sequential data which is becoming increasingly available as monitoring technologies are applied.

%We report an accuracy of 80\% in identifying the time series of those will receive a Alzheimer's disease, and demonstrate that the interaction between ROI volumes may be predictive.
\end{abstract}


\end{frontmatter}

%\linenumbers

%____________________________________________________________________________________________________
%
%   S E C T I O N
%____________________________________________________________________________________________________
% This section is the same as in Paper 1.
\section{Introduction}
Alzheimer's disease (AD) is an irreversible brain disorder which progressively affects cognition and behaviour, and results in an impairment in the ability to perform daily activities.  It is the most common form of dementia in older people, affecting about 6\% of the population aged over 65, and it increases in incidence with age. The initial stage of AD is characterised by memory loss, and this is the usual presenting symptom.  Memory loss is one constituent of mild cognitive impairment (MCI), which can be an early sign of Alzheimer's disease.  MCI is diagnosed by complaints of subjective memory loss (preferably corroborated by a close associate or partner of the individual), impairment of memory function, unimpaired general cognition and behaviour but with no evidence of dementia \cite{nestor2004}.  MCI does not always progress to dementia or to a diagnosis of Alzheimer's disease, but those with amnestic mild MCI, the type of MCI characterised by memory impairment, are more likely to develop dementia than those without this diagnosis.  In cases where an individual does develop Alzheimer's disease, the phase of MCI ends with a marked decline in cognitive function, lasting two to five years, in which semantic memory (the recall of facts and general knowledge) and implicit memory (the long-term, nonconscious memory evidenced by priming effects) also becomes degraded.  

% This section is the same as in Paper 1.
%____________________________________________________________________________________________________
%
%   S E C T I O N
%____________________________________________________________________________________________________
\section{Predicting Alzheimer's disease}

The disease pathology leads to a progressive, irreversible loss of brain function which suggests that prospective drug therapies should be tested for efficacy as early in the process as possible.  So there has been a demand for predicting which individuals will develop AD as early as possible in order to test drug therapies which might inhibit or prevent tissue damage. Input variables for use with learning methods may be derived from imaging (in particular MRI), cognitive tests, physical biomarkers such as APOE status and demographic variables such as age and gender.
\vspace{-4mm}
\paragraph{Data}
% See http://adni.loni.usc.edu/about/
There are a number of data repositories which are used for studies of Alzheimer's disease.  ADNI, the Alzheimer’s Disease Neuroimaging Initiative has been used in many studies\cite{weiner2017recent}. ADNI is comprised of four phases: ADNI-1 (2004), ADNI-GO (2009), ADNI-2 (2011), and ADNI-3 (2016).  ADNI is led by Principal Investigator Michael W. Weiner, MD. For up-to-date information see \url{www.adni-info.org}.  Another large scale study is AIBL, The Australian Imaging Biomarkers and Lifestyle flagship study of ageing which started in 2016\cite{ellis2009australian}. One European study is AddNeuroMed which is a public/private consortium developed for AD biomarker discovery \cite{lovestone2009addneuromed}, and it has been used for an number of comparative studies.  Of these, ADNI is the longest running and most cited study, although there are many other data sets and initiatives integrating together different datasets.
% Could add AddNeuroMed.

% ADNI-1 registered 200 healthy elderly, 400 participants with MCI, and 200 participants with AD, and the subsequent phases continued to add participants.  ADNI-1, began with 400 subjects diagnosed with mild cognitive impairment (MCI), 200 subjects with early AD and 200 elderly control subjects (labelled CN).  ADNI-1 was extended with ADNI GO which assessed the existing ADNI-1 cohort and added 200 participants identified as having early mild cognitive impairment (EMCI). As ADNI GO was ending, ADNI-2 began and assesses participants from the ADNI-1/ADNI GO cohort in addition to the following new participants: 150 elderly controls, 100 EMCI participants, 150 LMCI (late “mild cognitive impairment”) participants and 150 mild AD patients. The brain regions of interest used in this study are available from the ADNI data set. 
\vspace{-4mm}
\paragraph{Competitions}
Competitions for accurate prediction of diagnosis and biomarker measures have allowed comparisons between learning methods and feature choice.  The CADDementia challenge \cite{bron2015standardized} compares algorithms for multi-class classification of AD, MCI and controls based on structural MRI data.  The Kaggle Neuroimaging challenge \url{https://www.kaggle.com/c/mci-prediction} \cite{sarica2018editorial} was hosted on the Kaggle machine learning platform and used ADNI data.  The challenge involved a four-fold classification into AD, MCI, MCI converters to AD, and controls.  The 2017 TADPOLE grand challenge \url{https://tadpole.grand-challenge.org/} is currently taking place with  evaluation to be completed by January 2019. The TADPOLE challenge has the aim of predicting the onset of Alzheimer's disease using different modes of measurement, including demographic, physical and cognitive data.  The challenge is a three-fold diagnosis classification into AD, MCI and control groups, and the prediction of ADAS-13 score and normalised brain volume.  A preprocessed form of the ADNI data from the TADPOLE grand challenge \url{https://tadpole.grand-challenge.org/} was used in the preparation of this report\cite{tadpole}.  
\vspace{-4mm}
\paragraph{Imaging features}
Brain imaging methods can be used to derive features for predicting diagnosis either by analysis of the whole brain (voxel-based morphometry) or by deriving features from specific brain regions that are known to change during the course of Alzheimer's disease.  Using ADNI data, Schmitter \etalns\cite{schmitter2015evaluation} found that volume-based morphometry achieved at least as good an accuracy as voxel-based morphometry for typical classification tasks. S\o rensen \etal examined the differential diagnosis of AD and MCI using features derived from MRI and found that the most important MRI biomarkers were; hippocampal volume, ventricular volume, hippocampal texture, and parietal lobe thickness\cite{sorensen2017differential}.   Using training data from ADNI and AIBL, they were ranked as first place on the CADDementia challenge with a multi-class classification accuracy of 63\%.  Westman \etalns\cite{westman2011addneuromed} used the ADNI and AddNeuroMed data sets to examine 34 regional cortical thickness measures and 23 volume measures and found in both cohorts the most important features to be the entorhinal cortex, hippocampus and amygdala volumes. %they also tried prediction using one cohort as a training set and vice-versa. 

%A meta analysis by Seo \etal \cite{seo2017} of 32 imaging studies concluded that MRI on entorhinal cortex atrophy is comparable in prediction value to that of amyloid PET.  Prestia \etalns\cite{prestia2015} found that the highest predictive accuracy was achieved by combinations of amyloidosis and neurodegeneration biomarkers \cite{prestia2015}.

It should be noted that there is heterogeneity among patients diagnosed with Alzheimer's disease: Poulakis \etalns\cite{poulakis2018heterogeneous} identified five groups of patients by clustering AddNeuroMed and ADNI data on cortical and subcortical volume measures.  The groups were labelled as minimal atrophy, limbic-predominant, hippocampal-sparing, and two diffuse atrophy subtypes, and have different demographic, clinical and cognitive characteristics and different rates of cognitive decline.

% \item \cite{ferreira2017interactive} - Age and global brain atrophy were the most important variables in explaining variability in hippocampal volume. These variables were not only important themselves but also in interaction with gender, education, MMSE, and total intracranial volume.

%____________________________________________________________________________________________________
%
%   S E C T I O N
%____________________________________________________________________________________________________

\section{Experiment}
In this paper, we aim to demonstrate the application of the path signature by distinguishing Alzheimer's and healthy groups using MRI data as input to a classifier.  In doing so we investigate some of the nonlinear interactions between features, that is the relative changes between brain regions that occur with aging.  The novel aspect is the use of the path signature for generating features for classification.  The path signature was originally introduced by Chen \cite{chen1958integration} who applied it to piecewise smooth paths, and it was further developed by Lyons and others \cite{boedihardjo2016signature}\cite{hambly2010uniqueness}\cite{lyons2014rough}\cite{lyonssystem}\cite{xie2017learning}.  It is a systematic way of providing feature sets for multimodal data that can probe the nonlinear interactions in the data as an extension of linear features. Since all time-dependent interactions between variables are encoded, it likely to be useful for detecting unforeseen relationships between variables. Combined with Lasso regularisation, it can detect situations where the observed data has nonlinear information that significantly improves the inference.  Second order information has proved useful in some applications, for example in \cite{arribas2017signature} where the method was applied to modelling bipolar disorder and borderline personality disorder.  
%____________________________________________________________________________________________________
%
%   S E C T I O N
%____________________________________________________________________________________________________

\begin{figure}[t]
\centering
\includegraphics[trim = 0mm 0mm 0mm 0mm, clip, width=4cm]{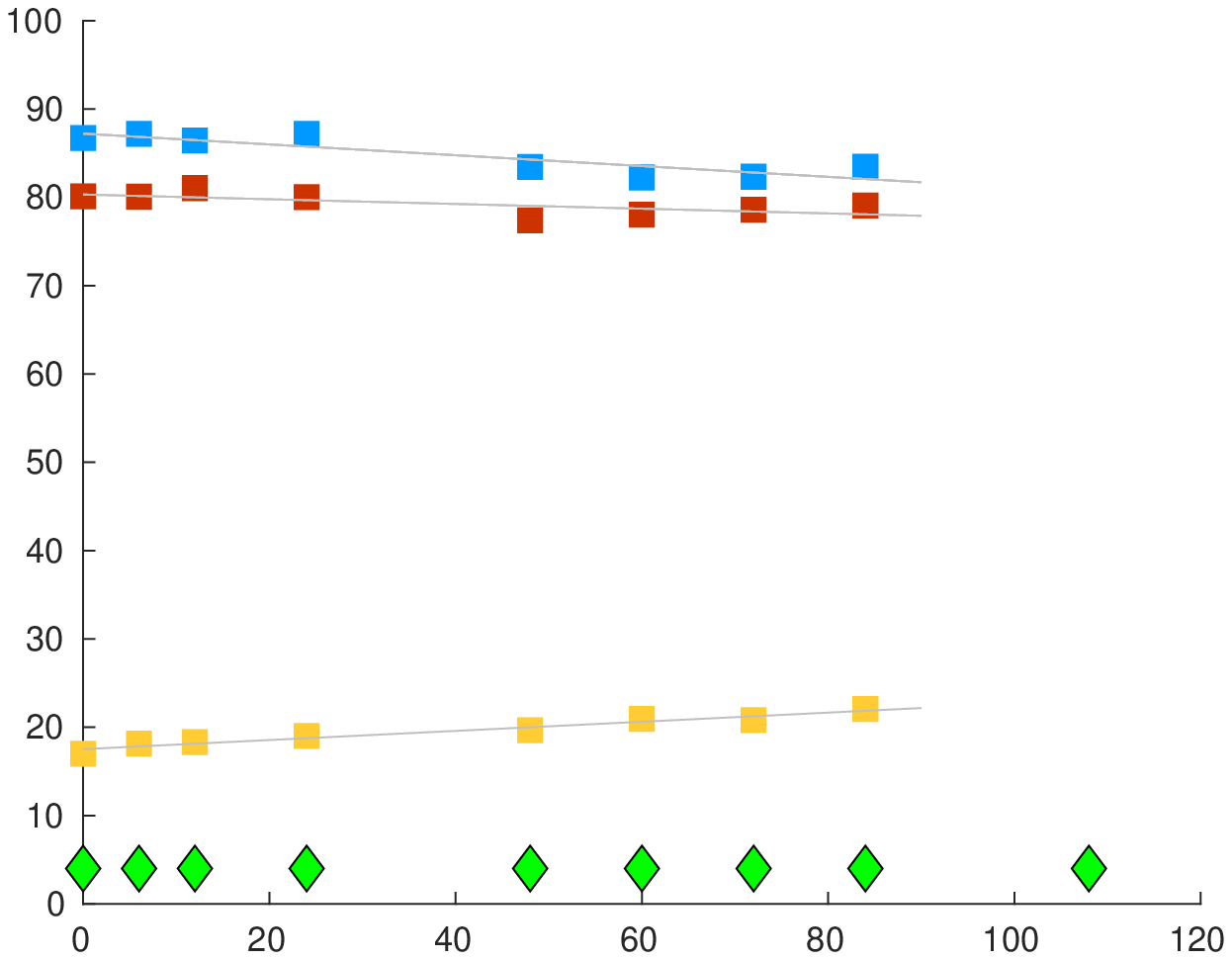} %left bottom right top
\includegraphics[trim = 0mm 0mm 0mm 0mm, clip, width=4cm]{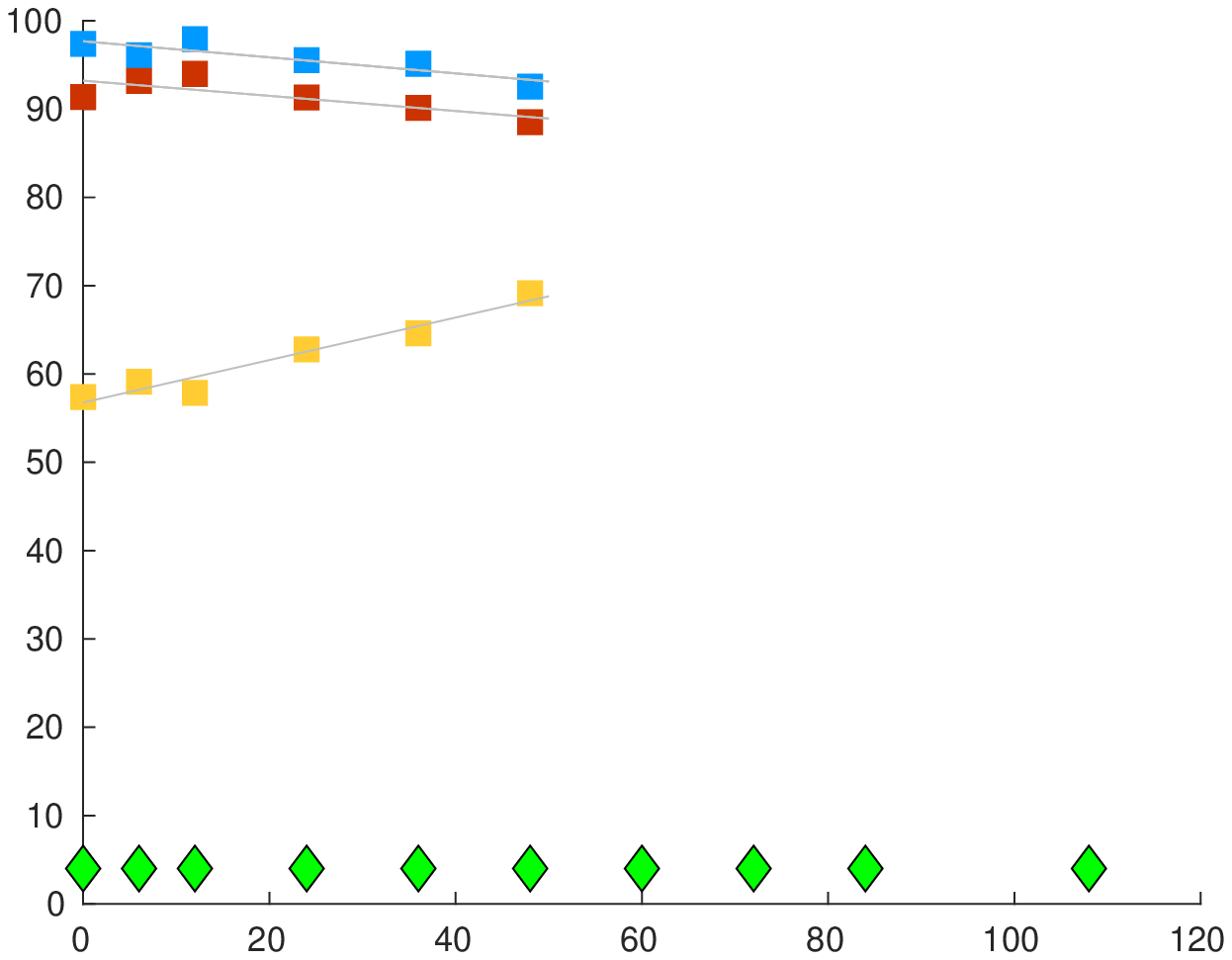} %left bottom right top
\includegraphics[trim = 0mm 0mm 0mm 0mm, clip, width=4cm]{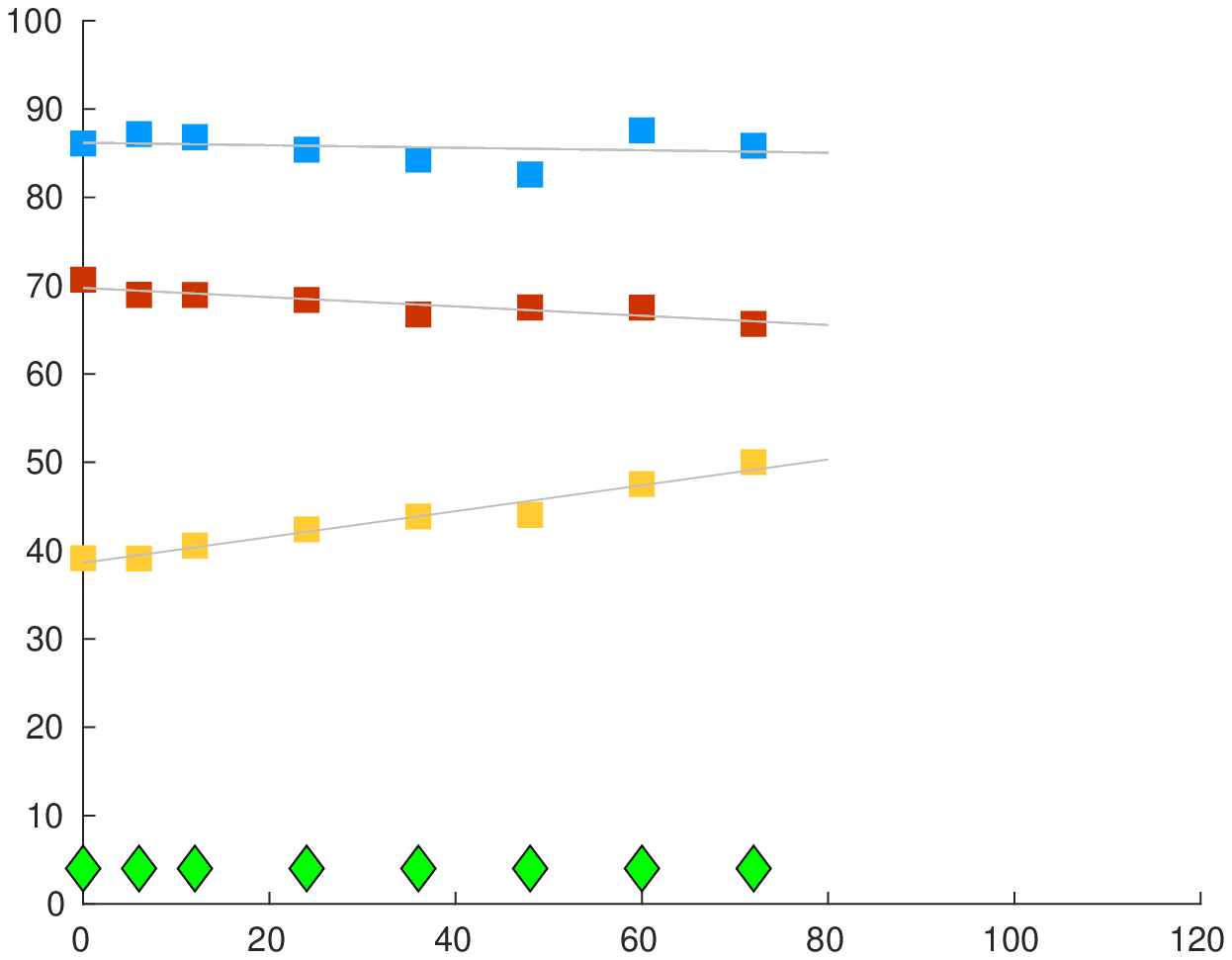} %left bottom right top
\includegraphics[trim = 0mm 0mm 0mm 0mm, clip, width=4cm]{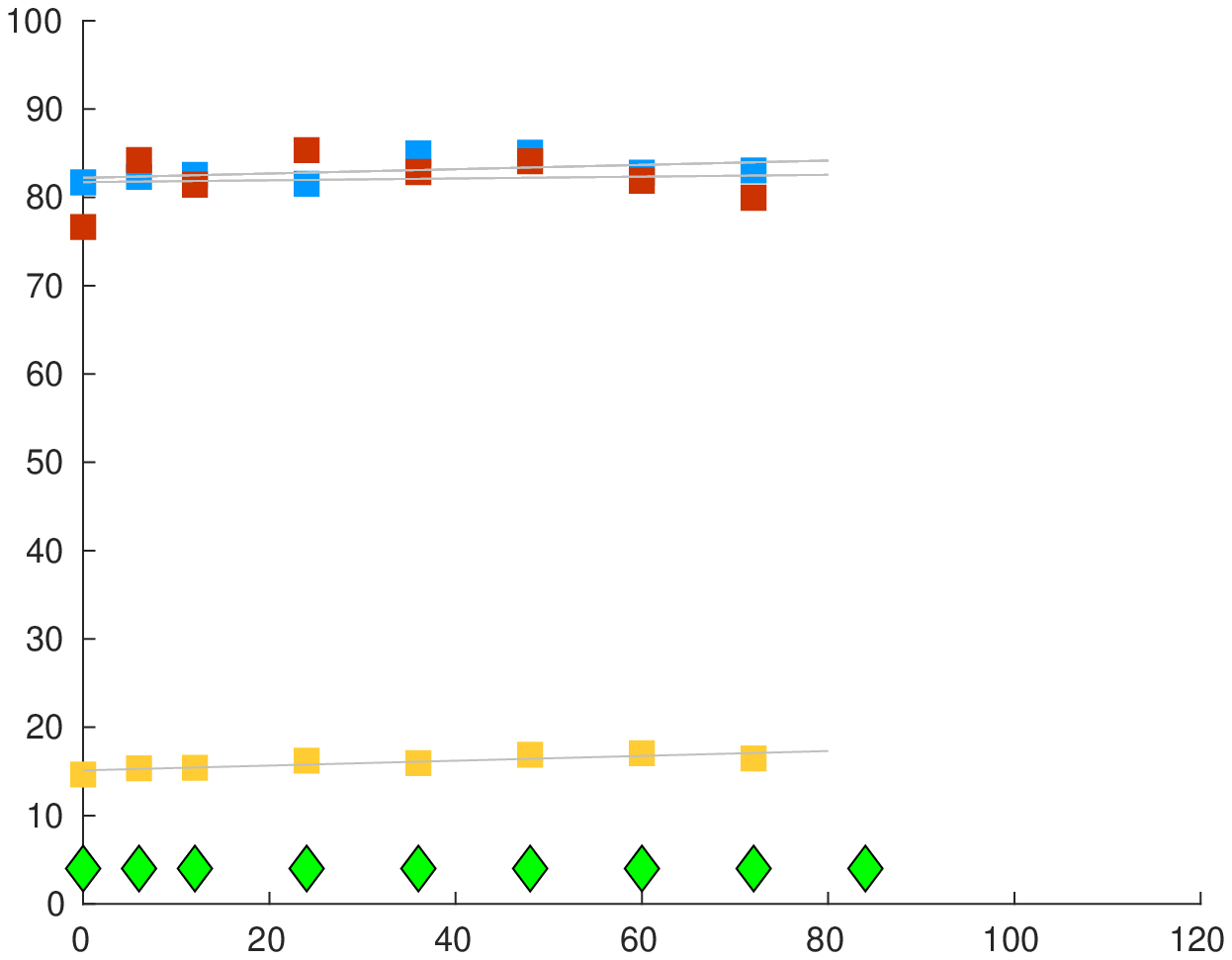} %left bottom right top
\\
\includegraphics[trim = 0mm 0mm 0mm 0mm, clip, width=4cm]{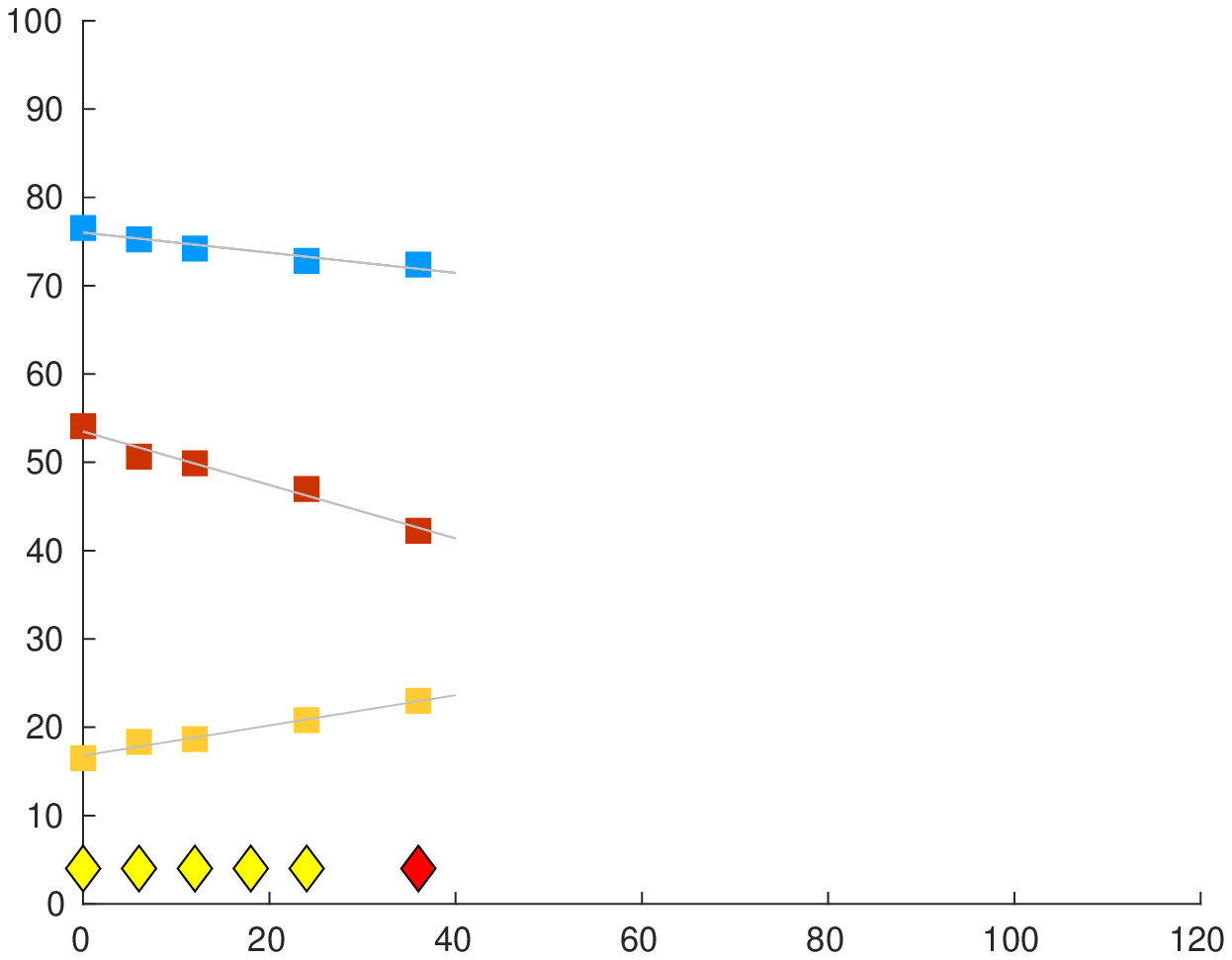} %left bottom right top
\includegraphics[trim = 0mm 0mm 0mm 0mm, clip, width=4cm]{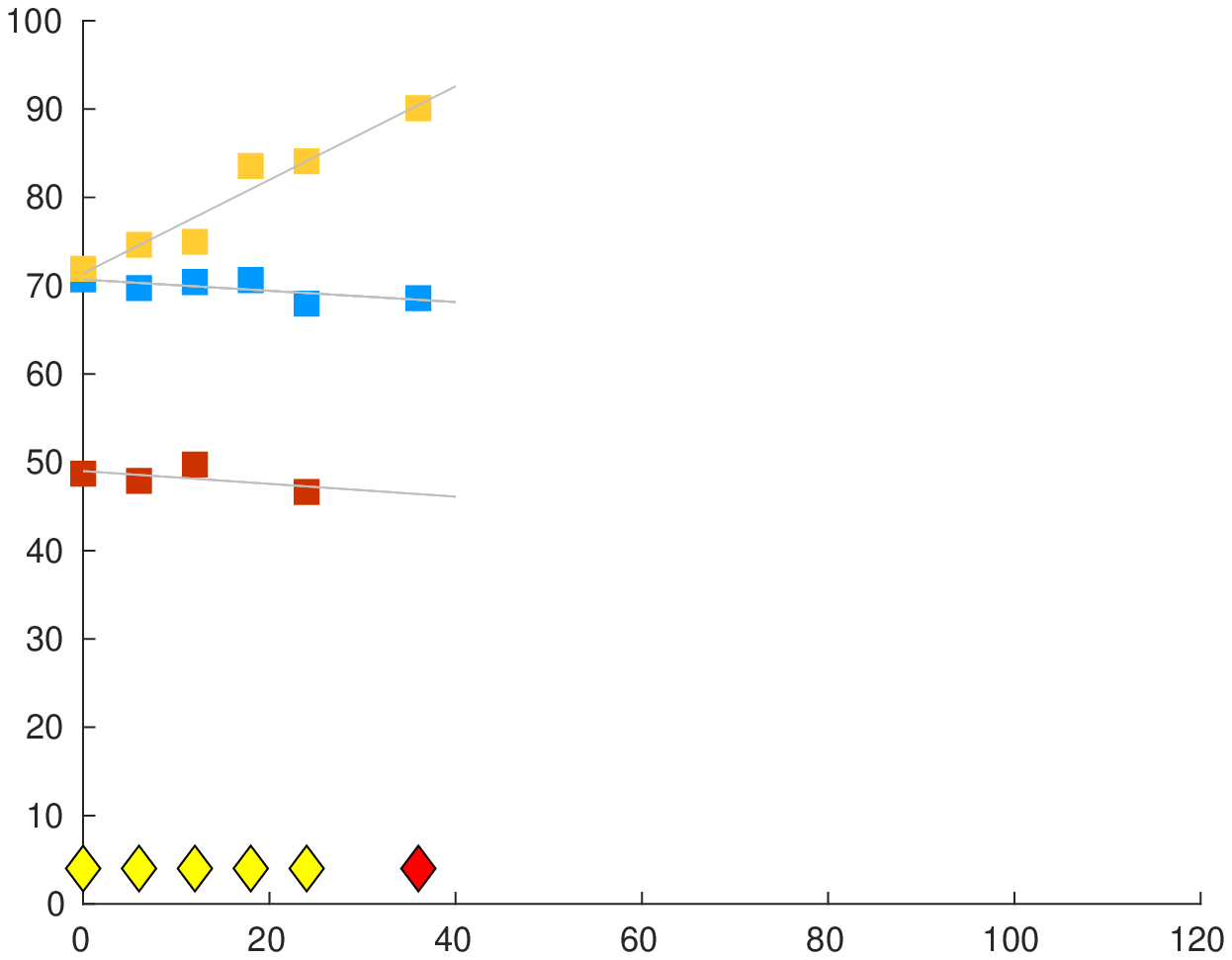} %left bottom right top
\includegraphics[trim = 0mm 0mm 0mm 0mm, clip, width=4cm]{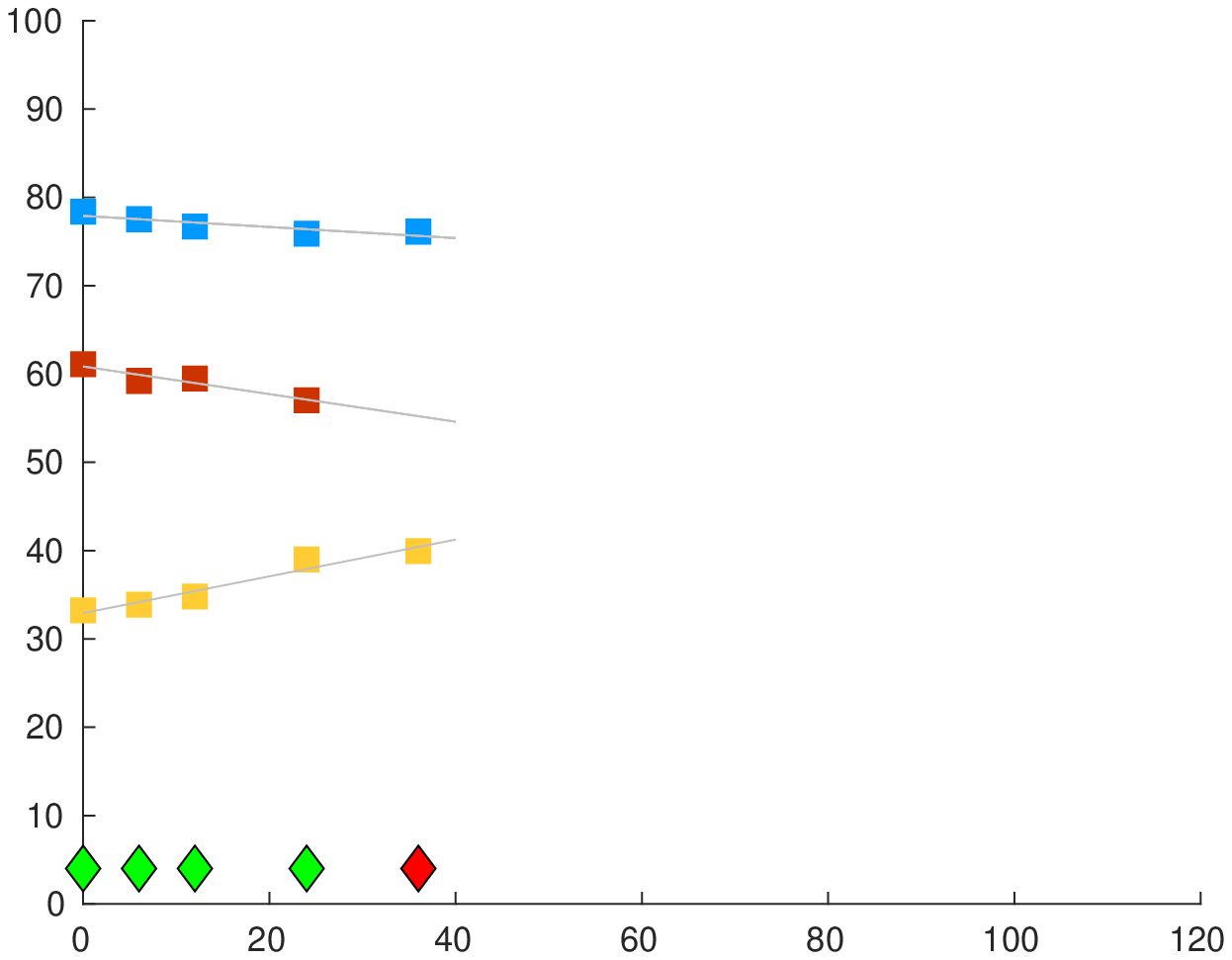} %left bottom right top
\includegraphics[trim = 0mm 0mm 0mm 0mm, clip, width=4cm]{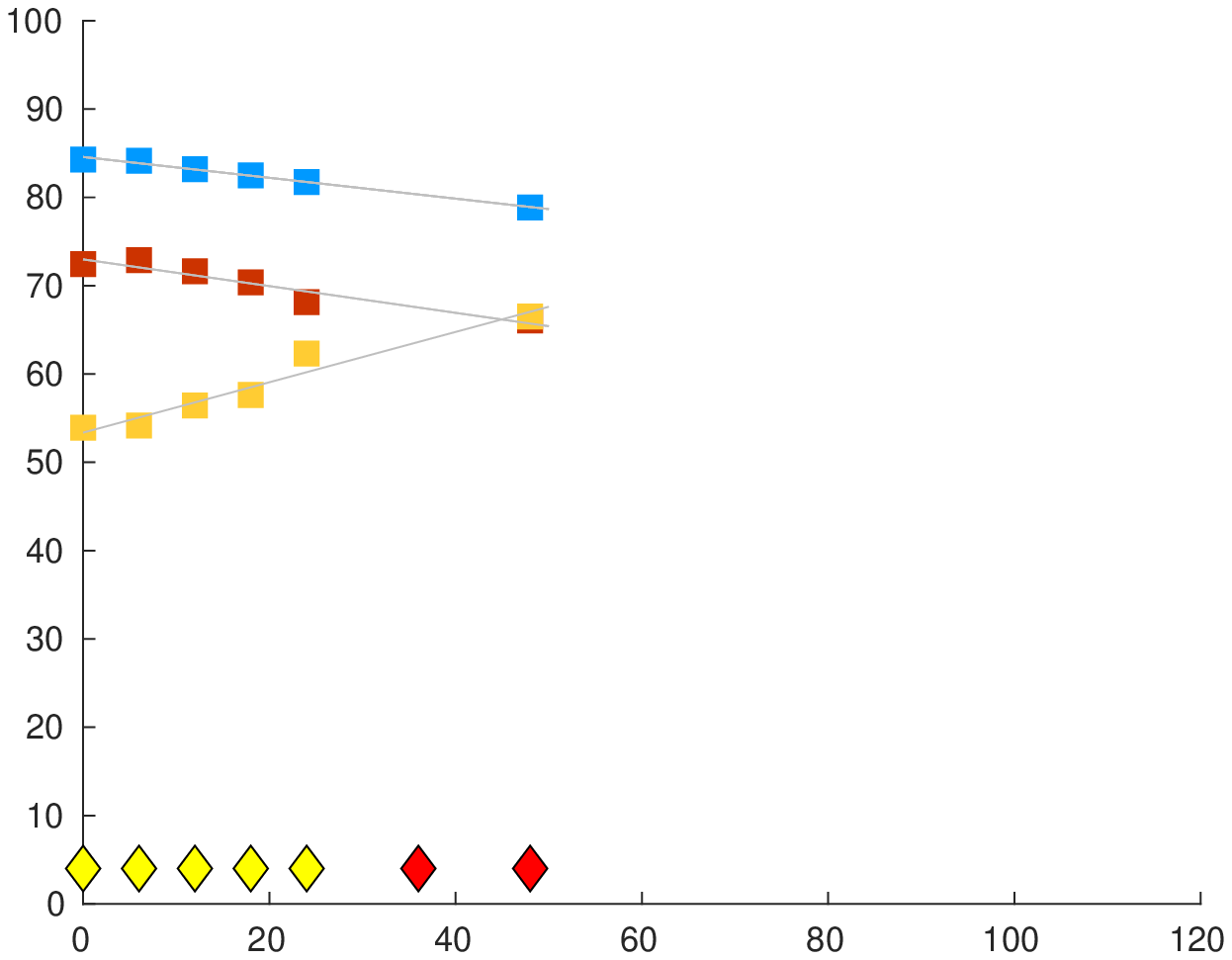} %left bottom right top
\caption[]{Sample plots by time (months) of scaled brain volumes for 8 patients, with the top four healthy and the bottom four those with Alzheimer's disease.   In each graph: Whole brain (blue), Hippocampus (red), Entorhinus (yellow).  Diagnosis points are diamonds: Healthy (green), MCI (yellow), Alzheimer's disease (red).}
\label{fig:rois}
\end{figure}

\vspace{-4mm}
\paragraph{Data}
We select two sets from the TADPOLE competition data for the first classification task.  A total of 1737 participants are first split into those who have a diagnosis of Alzheimer’s disease (AD) at some point (n=688), and the remainder who remain healthy (NL) (n=341).  The AD set is further selected to have the first diagnosis of AD at 36 months from baseline and with at least four measurements of all the variables WholeBrain, Hippocampus and Ventricles in the 24 months since baseline, with one measurement at the 24 months time point.  The participant must also have a match in the NL set.  To qualify to be a match to an AD participant, an individual must match their age to within 6 years and have healthy diagnoses for 72 months since the baseline measurement.  The matching time series must also have at least four measurement points up to month 24 including a measurement at month 24 itself.  Characteristics of the selected sets are shown in Table \ref{tab:set_comparison}, and sample plots of the time series are shown in Figure \ref{fig:rois}.

%____________________________________________________________________________________________________
%
%   T A B L E
%____________________________________________________________________________________________________
\begin{table}[htp]
\footnotesize
\newcolumntype{x}[1]{>{\raggedleft\arraybackslash\hspace{0pt}}p{#1}}
\newcolumntype{P}{p{20mm}}

\centering
\begin{tabularx}{0.5\textwidth}{x{25mm}PP}
 & \it{Healthy} &  \it{Alzheimer's} \\
\midrule
\it{n:} & 21 & 21 \\         
\\
\it{Age: min(mean)max} & 65.1 (74.8) 89.6 & 64.6 (75.4) 85.9  \\

%\it{min (mean) max:} & & \\
\\
\it{Gender M/F:} & 10/11 & 10/11 \\
\\
\it{Wholebrain: } & 1.05 (0.11)  & 1.00 (0.12) \\
\it{Hippocampus:} & 7.67 (0.84)  & 6.14 (1.41) \\
\it{Ventricles: } & 25.43 (19.91)  & 34.34 (21.17) \\
\it{median(iqr)}\\
\\
\it{APOE 0:} & 81\% & 29\% \\
\it{1:} & 19\% & 52\% \\
\it{2:} &  0\% & 19\% \\
\\
\it{MMSE 25} &  & 3 \\
\it{26} &  & 2 \\
\it{27} & 1 & 4 \\
\it{28} & 4 & 5 \\
\it{29} & 8 & 4 \\
\it{30} & 8 & 3 \\
\midrule
\end{tabularx}
\caption[]{Demographic and time series characteristics, and APOE4 status for matched healthy participants compared with selected participants who have a diagnosis of AD.  The variables shown are the sample size $n$, age, gender, scaled ROI volumes, APOE4 status and MMSE score at baseline.  In the AD set, 18 of the 21 participants have a diagnosis of MCI at baseline with the remaining 3 having a healthy diagnosis.}
\label{tab:set_comparison}
\end{table}
% source: create_time_matched_sets.m
%____________________________________________________________________________________________________
%
%   F I G U R E
%____________________________________________________________________________________________________                   

% source: plot_roi_time_series.m
%____________________________________________________________________________________________________
%
%   S E C T I O N
%____________________________________________________________________________________________________
% Delete We train a logistic regression model for classifying the time series.  
%For each time series a feature vector is derived from the measurement times and three brain volume measurements.  
\vspace{-4mm}
\paragraph{Method}
We take each time series from the AD set up to the last time point before diagnosis and extract a similar length time series from the counterpart in the NL set. The experiment is to identify time series as belonging to either the AD or NL set by using changes in relative brain volumes and time. For classification we use binary logistic regression which models the log probabilities of the outputs as linear functions of the inputs.  The inputs to the classifier are found using Lasso regularisation which subtracts an $L_1$ penalty from the negative log-likelihood when fitting the model.  The input to Lasso is 
a vector formed from the three ROI variables: WholeBrain, Hippocampus and Ventricles, and the path signature which is itself derived from these  variables and their time points.  The path signature maps the path of the input variables to a series of real numbers which uniquely characterise that path, $S = \{1, S^{(1)}, S^{(2)}, S^{(3)}, S^{(4)}, S^{(1,1)}, S^{(1,2)}, S^{(2,2)} , S^{(3,2)} , S^{(4,2)} \dots\}$.  The superscripts are multi-indexes or \emph{words} whose total length is in practice limited to degree $k$.  In this application we use $k=2$, resulting in a signature length of 21 where the first term, which is always 1, is unused in the feature vector.  The next four terms $S^{(1)}, S^{(2)}, S^{(3)}, S^{(4)}$ represent the increment of each variable, so in the experiment $S^{(1)}$ is the time period.  The next term $S^{(1,1)}$ is equal to $\frac{1}{2} (S^{(1,1)})^2$.  The cross terms $S^{(1,2)}$ and $S^{(2,1)}$ both represent the area between the actual path and a stepped path which increments one coordinate to its final value, then the next coordinate to its final value.   The log signature is an alternative representation that results from taking the logarithm in the formal power series and it is a more compact representation of the path than the signature. An introduction to the path signature which gives a geometric explanation of terms up to second  degree can be found in \cite{chevyrev2016primer}.  
\nind
Training is performed on the entire set of 42 time series using 10-fold cross-validation to find the graph of deviance against the Lasso coefficient $\lambda$ which determines the strength of regularisation.  As the Lasso coefficient increases, most of the variable coefficients shrink to zero, leaving a set of variables which act as predictors.

%Recent applications have been highly successful: they include Chinese text recognition \cite{xie2017learning} and human action recognition \cite{yang2017leveraging}. Path signatures provide a common analysis platform for both linear and nonlinear prediction and so can encode complex time ordered relationships between variables.  For example, the case where a cognitive faculty declines after cessation of smoking is distinguished from that when there is cognitive decline but smoking continues.  Since all time-dependent interactions between variables are encoded, it likely to be useful for detecting unforeseen relationships between variables.

%\vspace{4mm}
%____________________________________________________________________________________________________
%
%   S E C T I O N
%____________________________________________________________________________________________________
\section{Results}
\begin{figure}[htp]
\centering
\includegraphics[trim = 0mm 0mm 0mm 0mm, clip, width=7cm]{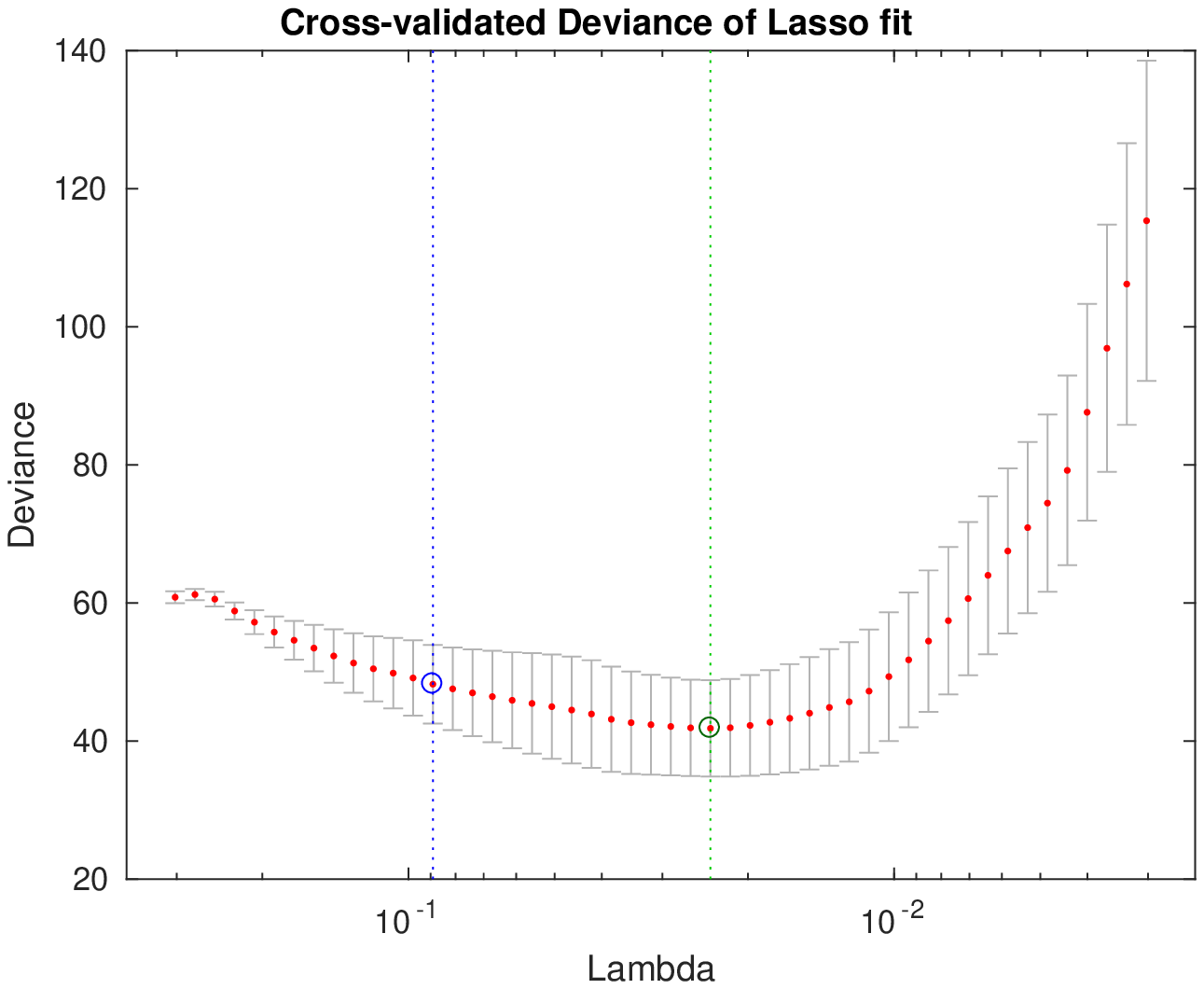} %left bottom right top
\includegraphics[trim = 0mm 0mm 0mm 0mm, clip, width=7cm]{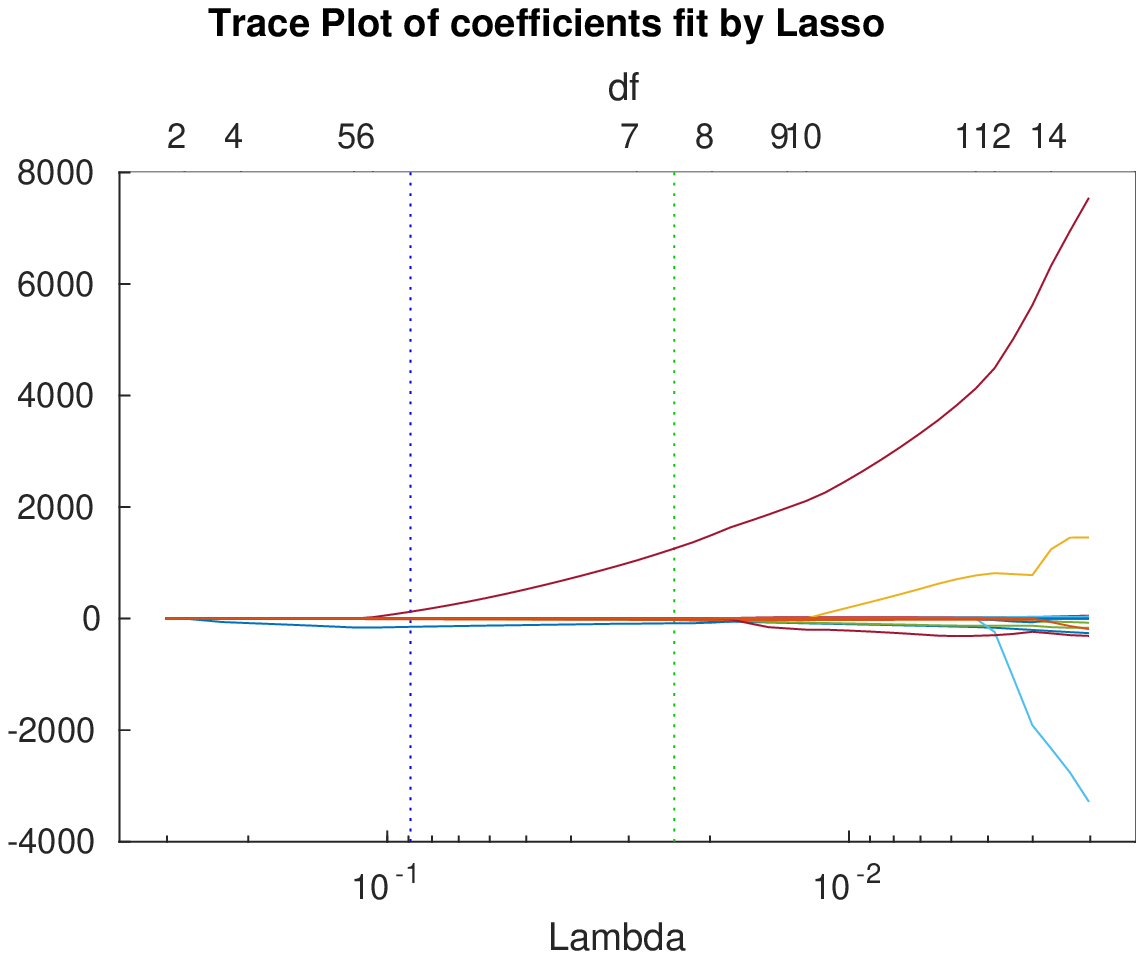} %left bottom right top
\caption[]{Signature training curves using baseline variable values and the path signature as features.  Top: Lasso parameter $\lambda$ optimisation using 10-fold cross validation.  The green circle and dashed line locate the Lambda with minimum cross-validation error. The blue circle and dashed line locate the point with minimum cross-validation error plus one standard deviation. Bottom: Coefficient shrinkage as $\lambda$ is increased. }
\label{fig:training_sig}
\end{figure}

\begin{figure}[t]
%\begin{figure}[htp]
\centering
\includegraphics[trim = 0mm 0mm 0mm 0mm, clip, width=7cm]{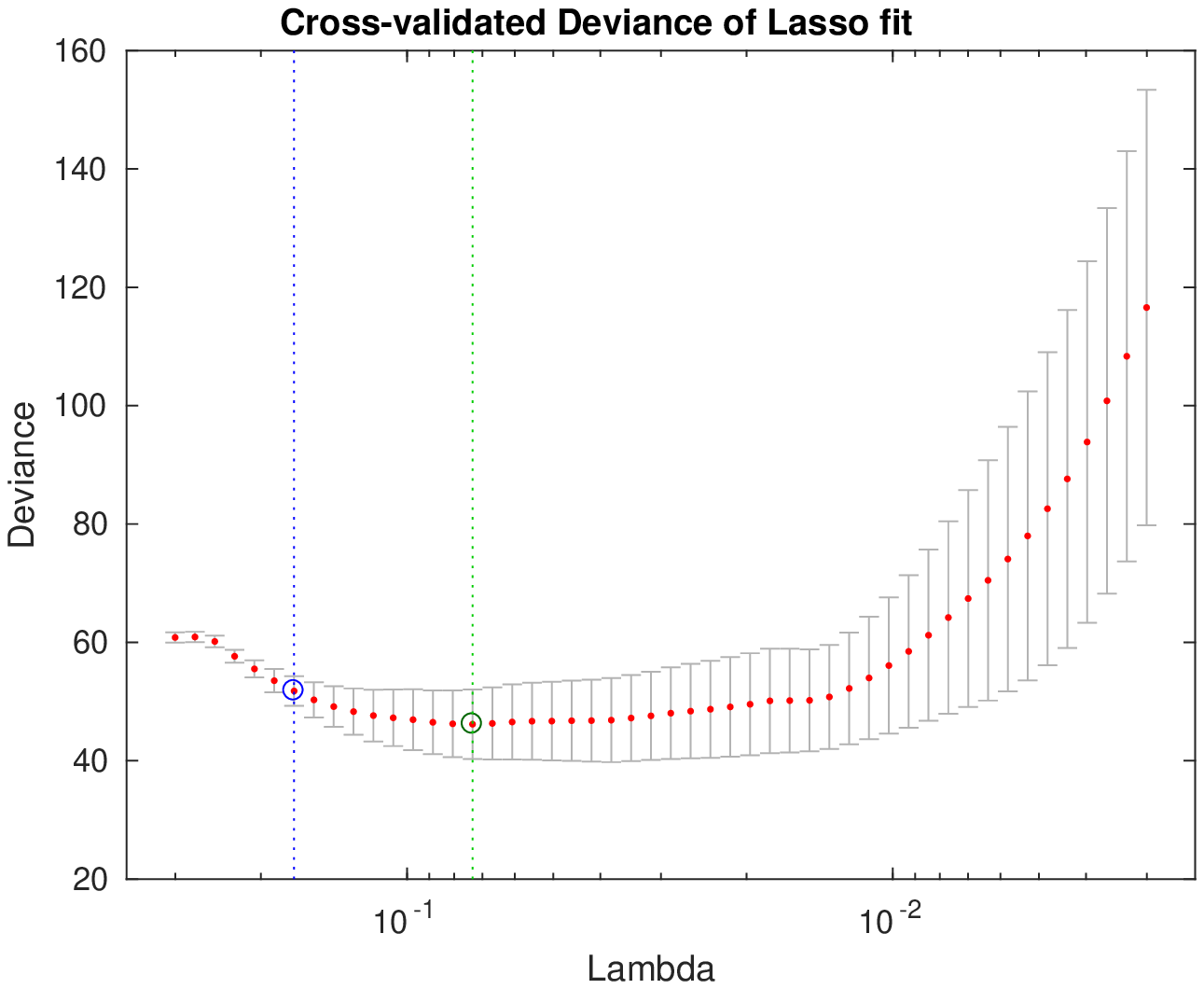} %left bottom right top
\includegraphics[trim = 0mm 0mm 0mm 0mm, clip, width=7cm]{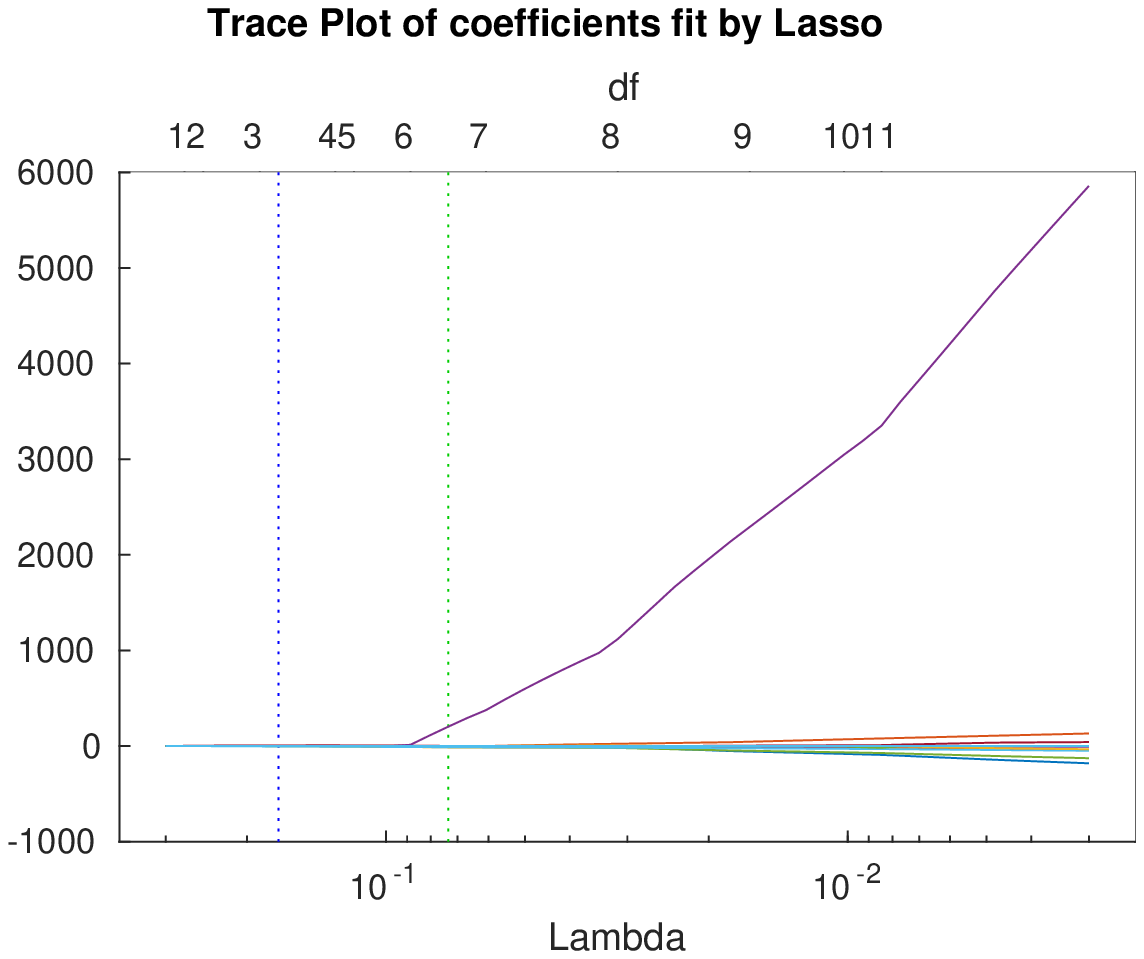} %left bottom right top
\caption[]{Log signature training curves using baseline variable values and the log path signature as features.   Top: Lasso parameter $\lambda$ optimisation using 10-fold cross validation.  The green circle and dashed line locate the Lambda with minimum cross-validation error. The blue circle and dashed line locate the point with minimum cross-validation error plus one standard deviation.  Bottom: Coefficient shrinkage as $\lambda$ is increased.}
\label{fig:training_logsig}
\end{figure}

Training curves for the signature and log signature features are shown in Figures \ref{fig:training_sig} and \ref{fig:training_logsig}. As lambda is increased (from right to left in the figures), the variable coefficients shrink and the deviance changes.   Table \ref{tab:selected_variables} shows those features that are selected at the point of minimum deviance plus one standard deviation. Using the log signature, Lasso selects the baseline (initial) volume of the ventricles and the change in volume of the hippocampus and ventricles.   Table \ref{tab:set_comparison} shows different ventricles sizes for the two sets, so it is not surprising that this variable was selected.  The selection of hippocampus and volume increments might also be expected from inspection of the sample graphs in \ref{fig:rois}.  Using the signature, Lasso selects some interaction terms.   The feature (t,t) represents the square of half the time increment: although the time series are all nominally of 24 months duration there are minor differences in the precise date when the MRI image was taken.  The remaining three features are interactions between variables (w,h), (w,v) and (h,t).   
% Delete The set of features selected in the case of the log signature feature set is simpler:  again the ventricles baseline is selected, and in addition the hippocampus and ventricle increments, showing that the reduction in size of these regions is predictive of a future diagnosis.
%Delete For the signature features, Lasso selects the baseline size of the ventricles, reflecting the wide variance of this value shown in Table \ref{tab:set_comparison}  
%____________________________________________________________________________________________________
%
%   T A B L E
%____________________________________________________________________________________________________

\begin{table}[htp]
\footnotesize
\newcolumntype{Y}{>{\centering\arraybackslash}X}
\centering
\begin{tabularx}{0.65\textwidth}{ll}
Signature & Log signature\\
\cmidrule(lr){1-2}

Ventricles baseline & Ventricles baseline \\
(t,t) & Hippocampus increment \\
(t,v) & Ventricles increment \\ 
(w,h) & \\
(w,v) & \\
(h,t) & \\
\cmidrule(lr){1-2}
\end{tabularx}
\caption[]{The set of variables selected by Lasso for the signature feature set shown in the left hand column, and the log signature feature set shown in the right hand column. The pairs $(t,t)$ etc. are nonlinear interactions between variables, with $(t,t)=0.5*incr(t)^2$ and for example $(t,v)$ representing the area between the variables Time and Ventricles.}
\label{tab:selected_variables}
\end{table}
% source: classify_time_matched_sets.m

%____________________________________________________________________________________________________
%
%   F I G U R E
%____________________________________________________________________________________________________

% source: classify_time_matched_sets.m

%____________________________________________________________________________________________________
%
%   F I G U R E
%____________________________________________________________________________________________________

% source: classify_time_matched_sets.m

%____________________________________________________________________________________________________
%
%   S E C T I O N
%____________________________________________________________________________________________________
%todo: update
\section{Conclusion}
The path signature has been used to encode the changes in time of brain ROI volumes for both healthy individuals and those with Alzheimer's disease.   We cannot at this stage draw strong conclusions from the selected set of variables beyond noting that they correspond to known Alzheimer's disease pathology, such as the changes in hippocampus and ventricles volume.  These processes also occur in normal aging, though to a lesser extent.  The identification of interaction terms formed of pairs of ROI variables is interesting and deserves further exploration.  The signature method generates interpretable nonlinear features and handles missing and irregular sequential data. While nonlinear classifiers (neural nets, random forests) can be accurate, their function can be difficult to interpret.  By encoding nonlinearity into the features, we can use classifiers that give more interpretable results.  Sequential data is becoming increasingly available as monitoring technology is applied, and the path signature method is a useful tool in the processing of this data.
%____________________________________________________________________________________________________
%
%   S E C T I O N
%____________________________________________________________________________________________________
%todo: update
\section{Acknowledgements}
Data collection and sharing for this project was funded by the Alzheimer's Disease Neuroimaging Initiative (ADNI) (National Institutes of Health Grant U01 AG024904) and
DOD ADNI (Department of Defense award number W81XWH-12-2-0012). A full list of funding sources for ADNI is provided in the document `Alzheimer’s Disease Neuroimaging Initiative (ADNI) Data Sharing and Publication Policy' available through \url{adni.loni.usc.edu/}.   
\nind
This work uses the TADPOLE data sets https://tadpole.grand-challenge.org constructed by the EuroPOND consortium http://europond.eu funded by the European Union’s Horizon 2020 research and innovation programme under grant agreement No 666992.  
\nind
The MRC Dementias Platform UK (DPUK) \url{https://www.dementiasplatform.uk/} provided support in the preparation of this paper.  DPUK is a multi-million pound public-private partnership, developed and led by the MRC, to accelerate progress in and open up dementias research. The aims of DPUK are early detection, improved treatment and ultimately the prevention of dementias.
%____________________________________________________________________________________________________
%
%   S E C T I O N
%____________________________________________________________________________________________________

\section*{}

\vspace{-5mm}
\bibliography{main}

\end{document}